\documentclass[aps,prb,reprint,superscriptaddress,floatfix]{revtex4-1}

\usepackage{graphicx,amssymb,bm,amsfonts,amsmath,graphics,color,epstopdf}
\usepackage[bookmarks=false,pdfstartview=FitH,colorlinks=true,citecolor=blue,linkcolor=blue]{hyperref}

\begin{document}

\title{Observation of Highly Dispersive Bands in Pure Thin Film C$_{60}$}

\author{Drew W. Latzke$^{\ddag}$}
\affiliation{Applied Science \& Technology, University of California, Berkeley, California 94720, USA}
\affiliation{Materials Sciences Division, Lawrence Berkeley National Laboratory, Berkeley, California 94720, USA}
\thanks{DWL and CO-A contributed equally to this work.}

\author{Claudia Ojeda-Aristizabal$^{\ddag}$}
\email[Corresponding author~]{Claudia.Ojeda-Aristizabal@csulb.edu}
\affiliation{Department of Physics and Astronomy, California State University, Long Beach, California 90840, USA}
\thanks{DWL and CO-A contributed equally to this work.}

\author{Sin\'{e}ad M. Griffin}
\affiliation{Molecular Foundry, Lawrence Berkeley National Laboratory, Berkeley, CA 94720, USA}
\affiliation{Department of Physics, University of California Berkeley, Berkeley, CA 94720, USA}

\author{Jonathan D. Denlinger}
\affiliation{Advanced Light Source, Lawrence Berkeley National Laboratory, Berkeley, California 94720, USA}

\author{Jeffrey B. Neaton}
\affiliation{Molecular Foundry, Lawrence Berkeley National Laboratory, Berkeley, CA 94720, USA}
\affiliation{Materials Sciences Division, Lawrence Berkeley National Laboratory, Berkeley, California 94720, USA}
\affiliation{Department of Physics, University of California, Berkeley, California 94720, USA}
\affiliation{Kavli Energy NanoSciences Institute at the University of California Berkeley and the Lawrence Berkeley National Laboratory, Berkeley, California 94720, USA}

\author{Alex Zettl}
\affiliation{Materials Sciences Division, Lawrence Berkeley National Laboratory, Berkeley, California 94720, USA}
\affiliation{Department of Physics, University of California, Berkeley, California 94720, USA}
\affiliation{Kavli Energy NanoSciences Institute at the University of California Berkeley and the Lawrence Berkeley National Laboratory, Berkeley, California 94720, USA}

\author{Alessandra Lanzara}
\email[Corresponding author~]{alanzara@lbl.gov}
\affiliation{Materials Sciences Division, Lawrence Berkeley National Laboratory, Berkeley, California 94720, USA}
\affiliation{Department of Physics, University of California, Berkeley, California 94720, USA}

\date{\today}

\begin{abstract}
While long-theorized, the direct observation of multiple highly-dispersive C$_{60}$ valence bands has eluded researchers for more than two decades due to a variety of intrinsic and extrinsic factors. 
Here we report the first realization of multiple highly-dispersive (330-520~meV) valence bands in pure thin film C$_{60}$ on a novel substrate---the three dimensional topological insulator Bi$_2$Se$_3$---through the use of angle-resolved photoemission spectroscopy (ARPES) and first-principles calculations. The effects of this novel substrate reducing C$_{60}$ rotational disorder are discussed. Our results provide important considerations for past and future band structure studies as well as the increasingly popular C$_{60}$ electronic device applications, especially those making use of heterostructures.
\end{abstract}


\maketitle

\section{\label{sec:Intro}Introduction}
C$_{60}$ has an unconventional zero-dimensional buckyball molecular structure that, when combined with its strong electron-electron and electron-phonon interactions \cite{Yang2003} in its bulk form, allow for unique properties not seen in ordinary (non-molecular) crystalline materials \cite{Brouet2006,Rao2010,Liao2014,Liu2014,Ojeda-Aristizabal2017,Kim2015,Cami2010,Sellgren2010,Campbell2015}. Bulk C$_{60}$ arranges itself in a face-centered cubic (fcc) lattice (one C$_{60}$ molecule centered at each lattice site), while in its thin film form it is deposited in layers corresponding to the (111) direction of the bulk lattice where each layer is arranged into a hexagonal lattice. As an initial approximation, one expects the electronic structure to be dominated by the electronic interactions within a single molecule---indeed the relative bond length between carbon atoms in a single molecule ($\sim$1~\AA) is much smaller than the bond length between the closest carbon atoms in adjacent molecules ($\sim$3~\AA) and the van der Waals bonds between adjacent C$_{60}$ molecules ($\sim$10~\AA) \cite{Wu1992}. However, whether such an approximation is valid is still widely debated. First principle calculations \cite{Ching1991,Saito1991,Troullier1992,Shirley1993} have reported relatively small bandwidths (0.5-1~eV) of bulk C$_{60}$ valence band manifolds but considerable band dispersion ($\sim$0.4-0.5~eV) of individual bands within those manifolds suggesting that it is inadequate to approximate the electronic structure of bulk C$_{60}$ with that of a simple isolated C$_{60}$ buckyball. In contrast, infrared and Raman spectroscopy studies have reported vibrational modes of solid C$_{60}$ consistent with a molecular solid \cite{Shinar2000}. Similarly, earlier photoemission spectroscopy studies have reported separate band manifolds with relatively small bandwidths ($<$1~eV) and small or unclear band dispersion \cite{He1995,He2007} (often only discernible at very low photon energies ($\lesssim$10~eV) \cite{Benning1994,Gensterblum1993}) for both the highest occupied molecular orbital (HOMO) and next highest occupied molecular orbital (HOMO-1) band manifolds. Whether such apparent diagreement is due to a combination of intrinsic and extrinsic factors, such as orientational disorder, transitions to excited vibrational states, electronic correlations, and/or finite resolution \cite{Wu1992,He2007,Louie1993} is still an open question.

Here we report the first observation of multiple highly dispersive bands in high quality C$_{60}$ thin films grown on a novel substrate, Bi$_2$Se$_3$, within the HOMO and HOMO-1 band manifolds using high-resolution angle-resolved photoemission spectroscopy (ARPES) measurements. These results are enabled by the excellent lattice match between the Bi$_2$Se$_3$ substrate and C$_{60}$ lattice and the constraints that the former imposes on the orientational order of the C$_{60}$ molecules. The agreement of our results with density functional theory (DFT) calculations supports the presence of a long range crystalline order in the C$_{60}$ thin films.

\section{Sample fabrication}
The high quality samples were grown on a bulk Bi$_2$Se$_3$ substrate cleaved in situ under ultra-high vacuum ($\sim$$10^{-10}$ Torr) before C$_{60}$ deposition using an effusion cell loaded with high purity (99.9\%) C$_{60}$ powder directed at the substrate. During the deposition (at $\sim$$1\times10^{-9}$~Torr), the sample was heated between 100-200$\,^{\circ}\mathrm{C}$ to facilitate the arrangement of large crystalline domains through increased C$_{60}$ mobility. A thickness of 5~nm (as measured by a quartz crystal thickness monitor) for the C$_{60}$ thin film was chosen to accurately probe the C$_{60}$ and minimize the substrate signal. We extract from our low-energy electron diffraction (LEED) measurements, shown in Fig.~\ref{fig:Fig1}(a), a nearest neighbor spacing of $\sim$10~\AA, similar to that of a bulk crystal. The clear LEED pattern consistent with C$_{60}$ structure and the well-matched lattice constants (see Supp.~Mat.~\cite{SupMatNote}) testify that Bi$_2$Se$_3$ is an excellent substrate for growth of high quality C$_{60}$ films. A further contributing factor to this harmony, as shown in Fig.~\ref{fig:Fig1}(c) and discussed later on, is discovered in our calculations which favor the hexagon faces of C$_{60}$, as opposed to the pentagon faces, to point towards the Se surface atoms.

\section{ARPES measurements}
High-resolution ARPES experiments were performed at Beamline 4.0.3 (MERLIN) of the Advanced Light Source using 45~eV linearly polarized (mostly out-of-plane) photons in a vacuum better than $5\times10^{-11}$~Torr. The total-energy resolution was 20~meV with an angular resolution ($\Delta\theta$) of $\leq0.2^{\circ}$. Data were taken at 20~K to assure the absence of spinning of the individual C$_{60}$ molecules, known to rotate and follow a ratcheting behavior above 50~K \cite{Dresselhaus1996}.

\begin{figure}
\includegraphics[width=\linewidth]{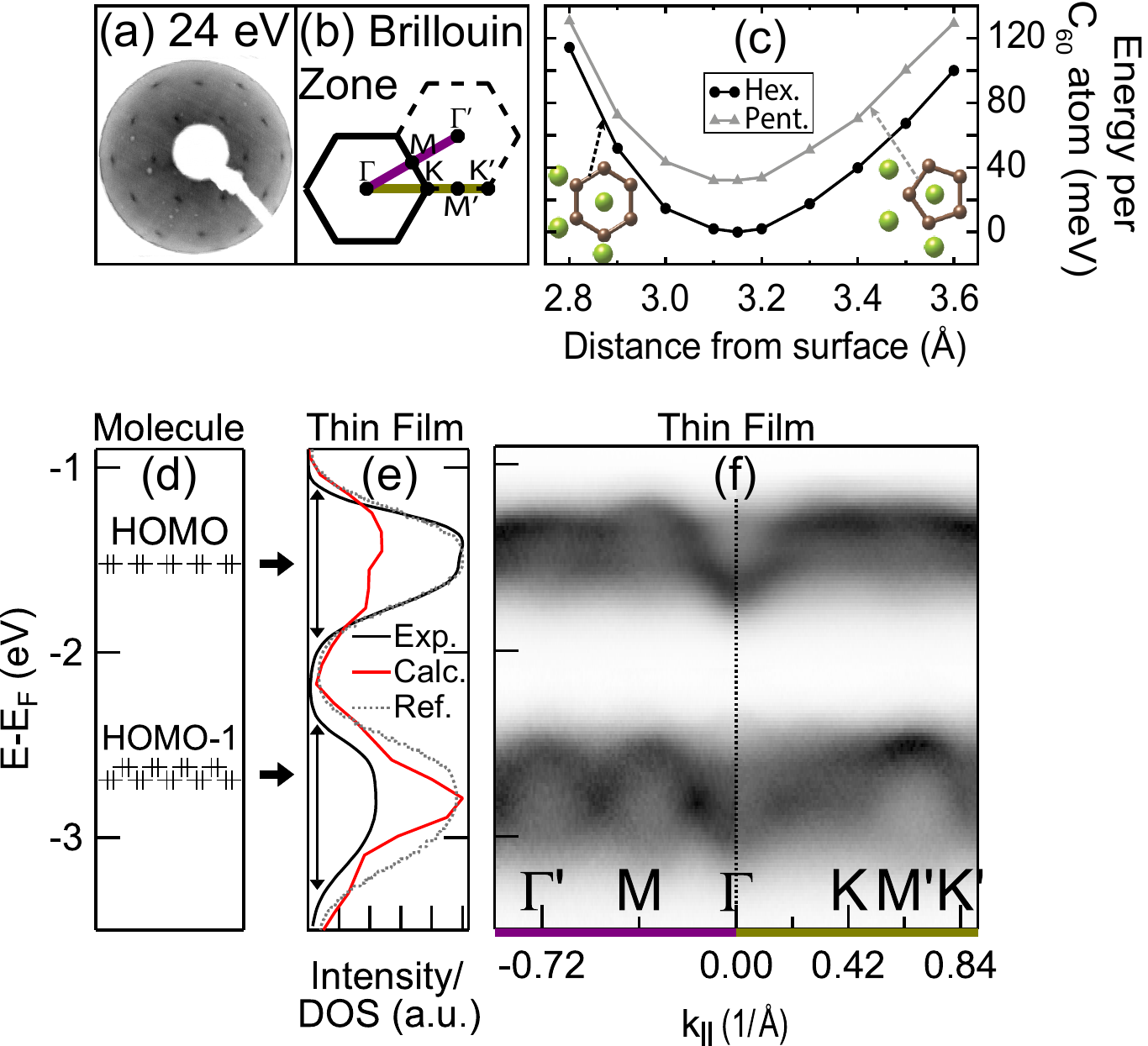}
\caption{\label{fig:Fig1}\textbf{(a)} LEED image of the crystalline 5~nm C$_{60}$ film. \textbf{(b)} Reduced surface Brillouin zone of thin film C$_{60}$. \textbf{(c)} Calculated DFT total energy for hexagon-down and pentagon-down C$_{60}$ molecules at various distances from a single layer of Se-terminated Bi$_{2}$Se$_{3}$ with van der Waals corrections included. Diagrams indicate the geometry of the bottom C$_{60}$ face (brown) on the substrate Se surface atoms (green). \textbf{(d)} Energy diagram comparing the molecular energy levels for the HOMO and HOMO-1 in a single C$_{60}$ molecule with \textbf{(e)} the corresponding band manifold in a thin film as shown by momentum-integrated ARPES intensity within the first Brillouin zone (solid black line), literature data \cite{Gibson2017} (dotted gray line), and the calculated density of states (DOS, calculation for single layer C$_{60}$, solid red line). \textbf{(f)} ARPES data along each of the high symmetry directions indicated in panel (b) showing large dispersions in the HOMO and HOMO-1 band manifolds. ($h\nu=45$ eV, $T=20$ K).}
\end{figure}

Fig.~\ref{fig:Fig1} gives an overview of the physical and electronic structure of the thin film (5~nm) C$_{60}$ sample. Panel (d) shows the HOMO and HOMO-1 energies for molecular C$_{60}$. These discrete energy levels evolve into dispersive manifolds for a crystalline lattice as a result of its symmetry and the interactions between C$_{60}$ molecules, as shown in panel (e). We see an overall good agreement between the momentum-integrated energy distribution curve (EDC) from our sample (black line) with the density of states (DOS, red line, computed with DFT for single layer C$_{60}$, see Supp. Mat. for details) and previous measurements on a different substrate (dotted gray line) \cite{Gibson2017} for both the HOMO and HOMO-1 centroid energies and bandwidths. 
The relative intensity of the two band manifolds is reversed when compared with the DFT DOS, but as shown in Supplementary Fig.~S3 and studied in detail elsewhere \cite{C60Paper2}, this is due to matrix element effects not considered in our calculations. The high quality of our C$_{60}$ thin film, enabled by epitaxial growth and a reduction of rotational disorder thanks to the Bi$_2$Se$_3$ substrate, allows us to resolve for the first time highly dispersive HOMO and HOMO-1 bands along the high symmetry directions, as shown in panel (f). The location of the cuts are shown in the Brillouin zone diagram in panel (b). 
 
\begin{figure}
\includegraphics[width=\linewidth]{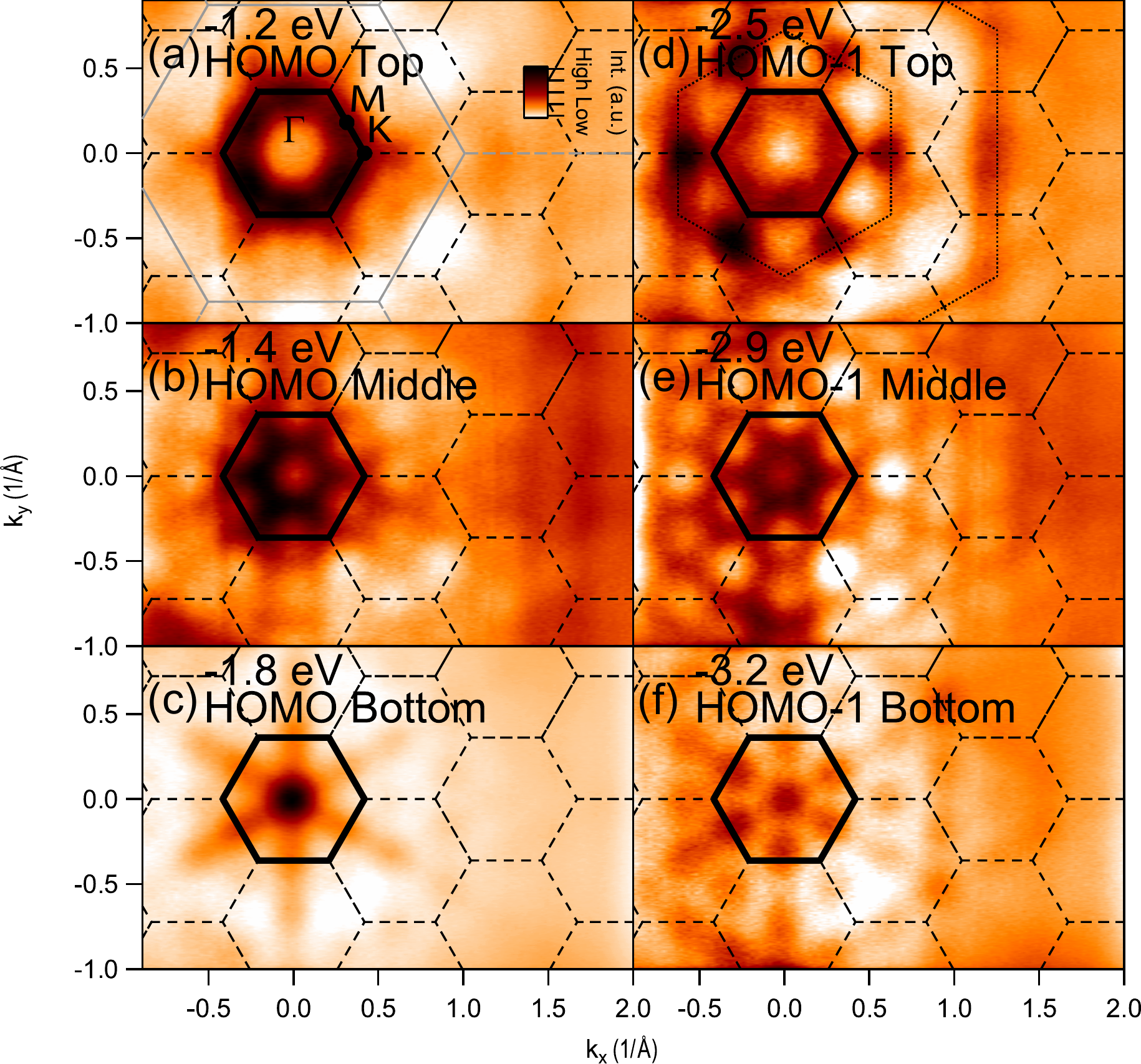}
\caption{\label{fig:Fig2}Constant energy maps of C$_{60}$ with energies near the \textbf{(a)} HOMO top, \textbf{(b)} middle, \textbf{(c)} bottom, and \textbf{(d)} HOMO-1 top, \textbf{(e)} middle, and \textbf{(f)} bottom. The first Brillouin zone of C$_{60}$ (Bi$_{2}$Se$_{3}$) is indicated by a thick black (gray) hexagon with the high-symmetry points labeled in panel (a). Dashed hexagons indicate higher order Brillouin zones, while two intensity patterns are highlighted in panel (d) by large dotted hexagons. ($h\nu=45$ eV, $T=20$ K).}
\end{figure}

Fig.~\ref{fig:Fig2} shows the momentum location and the energy dispersion of the HOMO and HOMO-1 bands over multiple Brillouin zones (dotted black hexagons). Brillouin zone size was calculated using a C$_{60}$ nearest neighbor spacing of 10.0~\AA, consistent with LEED measurements (Fig.~\ref{fig:Fig1}(a)) and our DFT calculations. Panels (a-f) show the constant energy maps at the top (a,d), middle (b,e), and bottom (c,f) of the HOMO and HOMO-1 band, respectively. In both cases, dipole matrix element effects enhance the intensity within the first Brillouin zone (thick black hexagon). 
For the top energy of the HOMO band manifold in panel (a), the most apparent feature is the strongly peaked hexagonal-like intensity pattern at the boundary of the first Brillouin zone. Moving to the middle energy of the HOMO band manifold (panel (b)), we see a highly periodic honeycomb-shaped structure which too has an enhanced intensity within the first Brillouin zone while faint highly periodic honeycomb-like features can still be observed in higher order Brillouin zones. These features decrease in size as we continue moving down in energy, eventually turning into an high intensity point at $\Gamma$ at the bottom of the HOMO band (panel (c)). Here in panel (c), we observe an additional enhancement of intensity along the $\Gamma-\text{M}$ high symmetry direction. The evolution of these features (panels (a)-(c)) is consistent with dispersive HOMO bands, whose minima occur at the $\Gamma$ and $\text{M}$ point.

The intensity maps of the HOMO-1 bands (Fig.~\ref{fig:Fig2}(d-f)) show strong similarities as well as peculiar differences with the HOMO dispersions at equivalent energies---the main difference being the redistribution of spectral weight within the high symmetry points and the different Brillouin zones. Within the first Brillouin zone, the top of the HOMO-1 band (panel (d)) shows the same hexagonal intensity pattern as the top of the HOMO band (panel (a)). At this energy we also observe two large hexagonal intensity patterns beyond the first zone, as marked by the dotted line hexagons, that evolve as binding energy increases. These hexagonal patterns appear to be rotated by 30$\,^{\circ}$ with respect to the first Brillouin zone boundary and are three and twelve times as large in size, respectively, with maximum spectral weight at discrete points along the $\Gamma-\text{K}$ direction.

As we continue decreasing in energy, the middle of the HOMO-1 (panel (e)) shows a honeycomb-shaped structure which decreases in size similarly as for the HOMO band, in agreement with an overall dispersion of the HOMO-1 band toward the $\Gamma$ and $\text{M}$ points. More specifically, we find the same honeycomb pattern found in the HOMO, but the effect of enhanced intensity is slightly reduced within the first Brillouin zone. Similarly, the bottom energy of the HOMO-1 (panel (f)) shows a similar pattern as the HOMO, but with the enhanced intensity effects reduced. A larger hexagonal pattern is again seen in the HOMO-1 at this energy.

While the origin of the hexagonal six-fold patterns is still not clear, when comparing them to the size and shape of the Bi$_{2}$Se$_{3}$ Brillouin zone (gray hexagon in panel (a)), we see that none of them align perfectly, making it unlikely that the substrate contributes significantly to the observation of these patterns. Additionally, comparisons with our DFT calculations (to be discussed later) suggest the direct influence of the substrate on the electronic structure of the C$_{60}$ film is negligible. In summary these energy maps reveal highly dispersive HOMO and HOMO-1 bands whose minima occur at the $\Gamma$ and $\text{M}$ points. The different intensity patterns between the two reveal different dipole matrix elements and likely small, but important, differences in orbital character. Further studies are needed to fully understand the underlying orbital characters of these bands including the precise details. The strong matrix element effects observed do not preclude the presence of a highly periodic band structure that spans beyond the first Brillouin zone (See Supp. Mat.).

Our findings underline the importance of taking into account matrix elements effects in C$_{60}$ and their effect beyond the first Brillouin zone. Indeed, many previous studies only considered dispersions near or within the first Brillouin zone \cite{Wertheim1995,Gensterblum1994,Tzeng2000,Gensterblum1993,Benning1994,He1995,Wu1992} and may have been susceptible to similar matrix elements effects that went unnoticed as the studies did not have a larger momentum range (like that of our study) to compare with. These effects could have inconspicuously affected their observations and conclusions. Additionally, with our relatively high photon energies ($\geq$30~eV), we are not susceptible to final state effects affecting the observed dispersions due to conduction band dispersion that may be present in previous studies claiming observation of HOMO and HOMO-1 dispersion using lower photon energies ($\lesssim$10~eV) \cite{Gensterblum1993,Benning1994}. 

\begin{figure}
\includegraphics[width=\linewidth]{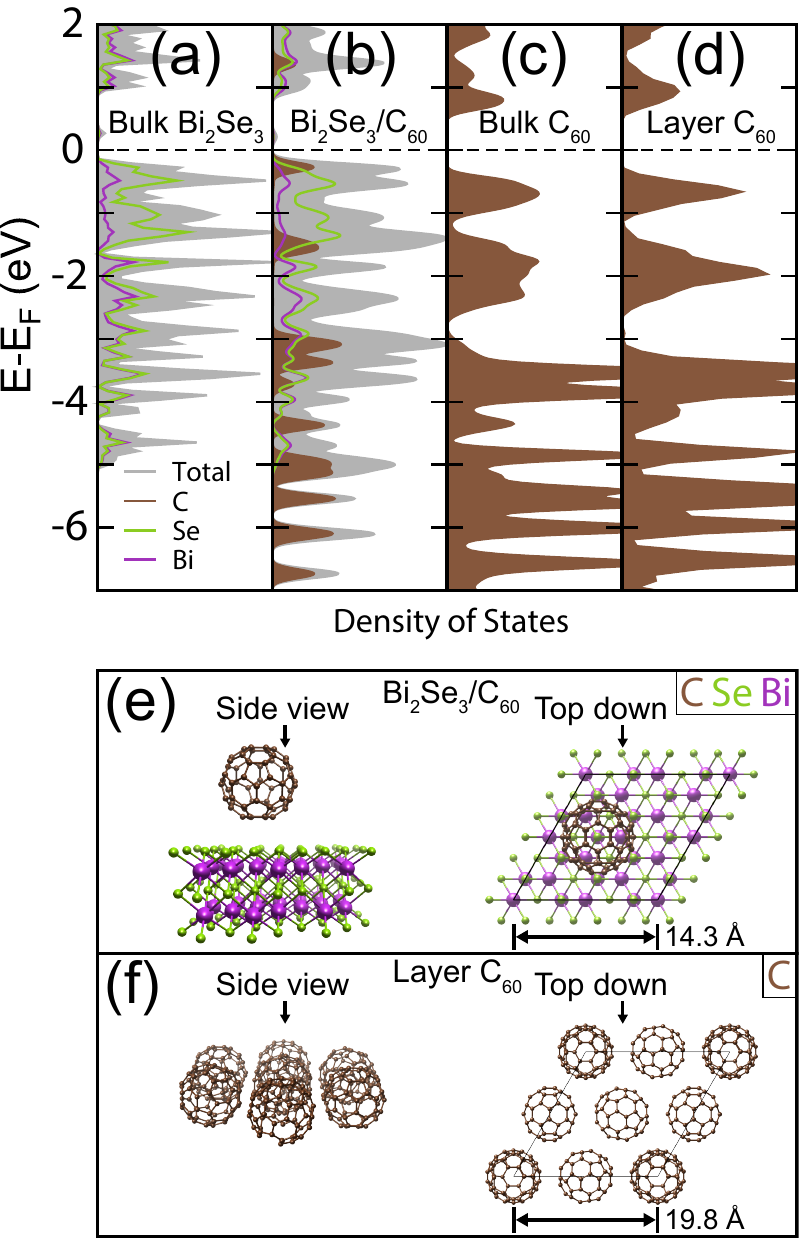}
\caption{\label{fig:Fig3}Calculated total and atom-projected density of states for \textbf{(a)} bulk Bi$_{2}$Se$_{3}$, \textbf{(b)} a single quintuple layer of Bi$_{2}$Se$_{3}$(0001) with a hexagon-down C$_{60}$ ball at its optimal distance of 3.15~\AA, \textbf{(c)} bulk C$_{60}$ in the $P\bar{a}3$ structure, and \textbf{(d)} a single hexagonal layer of C$_{60}$ in the (111) direction of the $P\bar{a}3$ structure. In each calculation both van der Waals corrections and the spin-orbit interaction were included. \textbf{(e)} Configuration of the Bi$_{2}$Se$_{3}$/C$_{60}$ structure used for DOS calculations in panel (b). \textbf{(f)} Configuration of the layer C$_{60}$ structure used for DOS calculations in panel (d). See text for further details.}
\end{figure}

\section{DFT calculations}
Figure \ref{fig:Fig3} shows our DFT calculations for C$_{60}$ and Bi$_{2}$Se$_{3}$ structures (see also Supp.~Mat.). We consider a representative interface of a C$_{60}$ monolayer with hexagonal symmetry on one quintuple layer of Bi$_2$Se$_3$(0001), where the separation between adjacent C$_{60}$ molecules is 14.3~\AA (in contrast to the experimental C$_{60}$ separation of 10A). The expanded lattice constant is used to diminish the computational demands of a full ab-initio calculation of the Bi$_2$Se$_3$/C$_{60}$ interface. We use this interface structure to determine the degree of electronic hybridization between the C$_{60}$ and substrate Bi$_2$Se$_3$ only, but do not (as will be elaborated) otherwise use it to compare to our experimental results. We calculated the relative energies of interfaces with Bi- and Se-terminated Bi$_{2}$Se$_{3}$ with both hexagon-down and pentagon-down C$_{60}$ for variety of C$_{60}$ distances from the substrate, as shown in Supplemental Fig.~S5. Our calculations revealed that the lowest-energy interface structure was a Se-terminated hexagon-down interface, and we refer to this interface structure for our remaining calculations. We found that the Se-termination is almost 1.4~eV per Se atom lower in energy than the Bi-termination using DFT plus van der Waals corrections \cite{Grimme_et_al:2010}, while the pentagon and hexagon C$_{60}$ interfaces are essentially degenerate when van der Waals corrections are not included. For the Se-terminated surface, we calculated the optimum distance of the C$_{60}$ monolayer from the surface. The hexagon-down geometry is lower in energy than the pentagon-down case by 32~meV per C$_{60}$ with an optimal distance of 3.15~\AA\ with PBE+vdW and 4.05~\AA\ for PBE as previously shown in Fig.~\ref{fig:Fig1}(c). Furthermore, we find a slight (few meV) preference for the hexagon to align along the hexagonal directions set by the Se atoms. The lower energy for the alignment of a C$_{60}$ hexagon face towards a Se atom, as compared with a pentagon face, imposes a constraint on the orientation of the C$_{60}$ molecules on the Bi$_{2}$Se$_{3}$, which may favor some long range order beneficial for ARPES experiments.

\begin{figure}
\includegraphics[width=\linewidth]{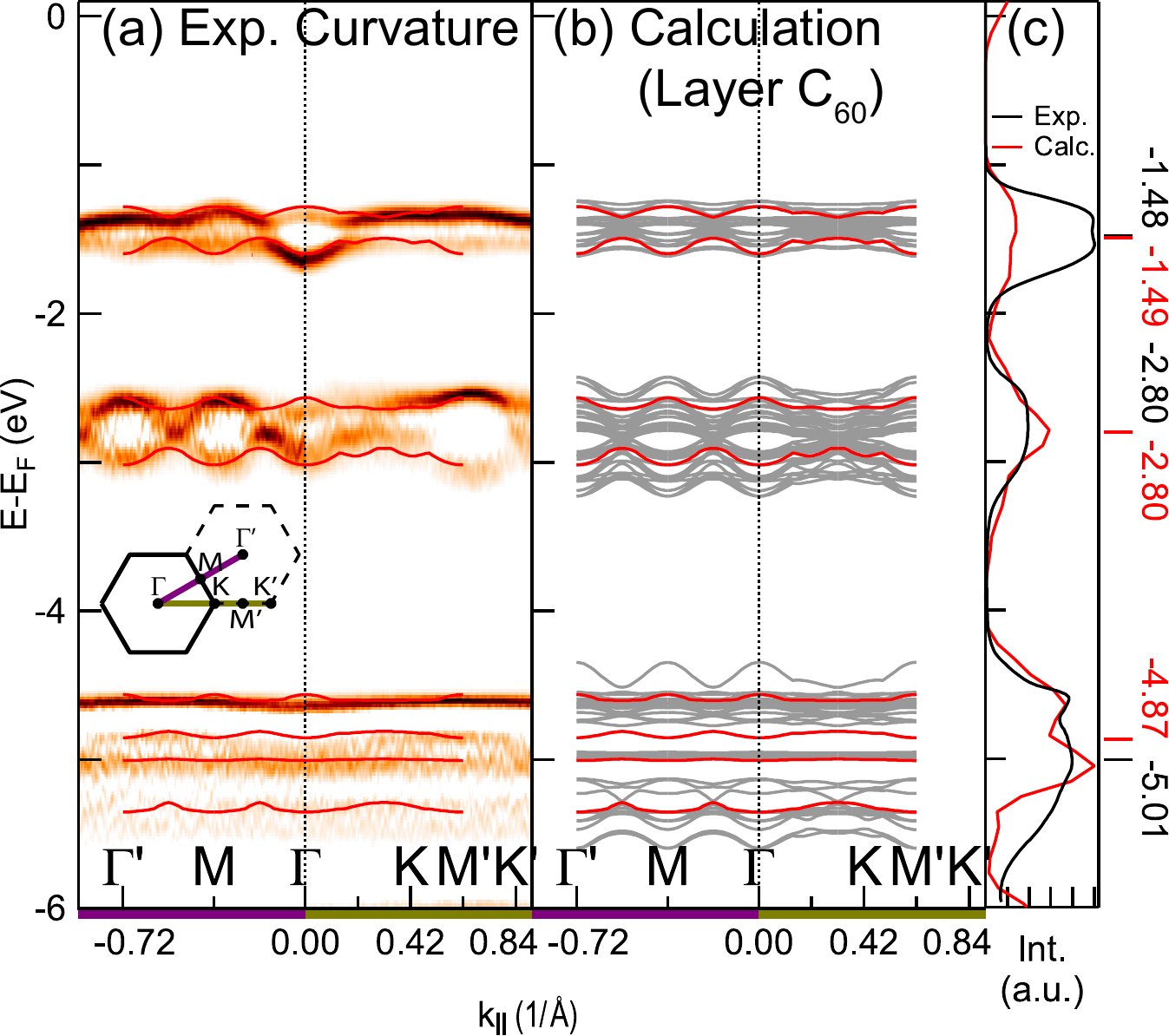}
\caption{\label{fig:Fig4}\textbf{(a)} Curvature of the electronic band structure of C$_{60}$ along the $\Gamma-\text{M}$ (purple) and $\Gamma-\text{K}$ (olive) high-symmetry directions. The calculated bands that best fit the HOMO, HOMO-1, and HOMO-2 experimental dispersions are plotted over the data as red lines. \textbf{(b)} Full DFT-calculated theory band structure of single layer C$_{60}$ including the same bands plotted in panel (a) highlighted in red. \textbf{(c)} Integrated EDC (black) across $\Gamma'-\text{M}-\Gamma-\text{M}-\Gamma'$ in comparison with DFT-calculated single layer C$_{60}$ density of states (red) indicating the centroid energy of each experimental and calculated band manifold peak. ($h\nu=45$ eV, $T=20$ K).}
\end{figure}

The calculated electronic density of states for C$_{60}$, Bi$_{2}$Se$_{3}$, and the interface structure are shown in Fig.~\ref{fig:Fig3}(a-d). Panel (a) shows the calculated density of states for bulk Bi$_{2}$Se$_{3}$ and panel (b) shows the lowest-energy Bi$_{2}$Se$_{3}$/C$_{60}$ interface structure. A representation of the interface structure is shown in panel (e). The interface DOS shows little hybridization between the Bi$_{2}$Se$_{3}$ substrate and C$_{60}$ molecule---the Bi and Se projected density of states in the Bi$_{2}$Se$_{3}$/C$_{60}$ interface are very similar to those in bulk Bi$_{2}$Se$_{3}$. Owing to the lack of hybridization between the Bi$_{2}$Se$_{3}$ and C$_{60}$ in our interface calculation (panel (b)), we calculate the density of states of bulk C$_{60}$ in the $P\bar{a}3$ structure (the calculated ground-state structure, panel (c)) and the density of states of a single hexagonal layer of C$_{60}$ (a slice of the $P\bar{a}3$ structure along (111), panel (d)). Interestingly, the density of states is quite similar in both cases suggesting that already in a single (111)-oriented layer, the C$_{60}$ band dispersion is quite robust, which is also is supported by our experiments. As a result of these conclusions, we use a single layer C$_{60}$ (structure shown in panel (f)) band structure calculations to compare with our experimental results, as will be discussed in Fig.~\ref{fig:Fig4}.

Following the lack of hybridization between states in the Bi$_2$Se$_3$ substrate and the C$_{60}$ thin film (concluded from our DFT results), in Figure \ref{fig:Fig4} we compare the experimentally measured valence bands of the C$_{60}$ thin film to the calculated band structure for a single (111)-oriented layer of C$_{60}$. Calculations are based this time on the 10~\AA\ nearest neighbor distance found in our LEED measurements. Fig.~\ref{fig:Fig4}(a) shows energy vs.~momentum cuts of the experimental thin film C$_{60}$ band structure along the $\Gamma-\text{M}$ and $\Gamma-\text{K}$ high-symmetry directions. The curvature in the energy dimension of the raw data (Fig.~\ref{fig:Fig1}(f)) is presented to more precisely locate and resolve the individual band dispersions. Two clear main dispersions are observed concurrently within each of the HOMO and HOMO-1 band manifolds which have bandwidths of 330~meV and 520~meV, respectively, based on the extent of the dispersions. These dispersion bandwidth values are $\sim$0.3 eV smaller than the full-width half-maximum (FWHM) spectral bandwidths, 0.62-0.66 eV and 0.80-0.83 eV, reported in previous photoemission studies  \cite{Gensterblum1993,Benning1991, He2007, Wang2008a} which include additional line-width broadenings.  See supplemental Fig. S7 for further discussion. 
Additionally, as discussed previously in Fig.~\ref{fig:Fig2}, for each band manifold these two dispersions provide both a local band maximum and minimum at $\Gamma$ (as well as at $\text{M}$), unveiling bands previously not resolved experimentally.

The high degree of similarity between the HOMO and HOMO-1 dispersions as noted in our constant energy maps (Fig.~\ref{fig:Fig2}) is again readily apparent. Similarities between angle-resolved EDCs for the HOMO and the HOMO-1 band manifolds were previously reported for studies on cleaved (111) surfaces of C$_{60}$ single crystals \cite{He1995} including the observation of a band minimum at $\Gamma$ in each of the band manifolds, which is compatible with our observations (in a thin film sample). In contrast, other studies have reported a local band maximum for the HOMO at $\Gamma$ \cite{Gensterblum1993}. Our measurements reconcile this apparent disagreement, while providing a full view of the dispersion of multiple bands within each band manifold and simultaneously resolving two bands in both the HOMO and HOMO-1 with a local band maximum and minimum at $\Gamma$. Additionally, we have resolved another valence band manifold (HOMO-2) at higher binding energy ($\sim$$4.5$~eV) where multiple weakly dispersive bands can be discerned.

Fig.~\ref{fig:Fig4}(b) shows the complete DFT-calculated band structure for single layer C$_{60}$ over the first three valence band manifolds. Despite lacking self-energy corrections, e.g.~within the GW approximation, the DFT spectra show an overall good agreement with our experimental data across multiple valence band manifolds. The calculated bands show an upper and lower grouping for both the HOMO and HOMO-1 that follows the path of the experimentally observed dispersions with singular bands that best fit the dispersions highlighted here in red (also plotted in panel (a) for direct comparison). Similarly, the HOMO-2 band manifold at high energy shows an excellent agreement with the theory calculations. The upper and lower groupings are easily observed as shoulders in the integrated EDC in Fig.~\ref{fig:Fig4}(c) (black). We find that these shoulders are not as well defined for lower quality samples lacking dispersive bands. The comparison of our experimental EDC with the theory-calculated density of states (red) shows an excellent agreement with only a 14.9\% expansion in the energy dimension of the theory data. The center energy for the HOMO and HOMO-1 fit within just 10~meV of the calculated band structure and the HOMO-2 within 140~meV.

The comparison of observed experimental dispersions and DFT results motivates a discussion on many body effects in C$_{60}$. It is known that electron correlations and electron-phonon coupling play an important role in the electronic properties of C$_{60}$ \cite{Knupfer1993,Yang2003,Brouet2006,Benning1993,Benning1993a,Goldoni2006}. The reported values for the Hubbard parameter U (on-site Coulomb interaction) ranges between 0.8~eV and 1.6~eV \cite{Lof1992,Golden1995,Dresselhaus1996,Antropov1992}, which is greater than the electronic bandwidth measured here (0.33~eV for the HOMO). This points towards strong electronic correlations in our thin film C$_{60}$, similar to what has been reported in the past for bulk C$_{60}$. 

\section{Conclusion}
In conclusion, we have identified a novel substrate for the growth of high quality thin film C$_{60}$, the topological insulator Bi$_2$Se$_3$. The constraints that this substrate imposes on the orientation of the buckyballs and its excellent lattice matching, support a long range crystalline order in C$_{60}$, enabling the first observation of a highly dispersive valence band structure, previously obscured by sample quality, momentum and energy resolution limitations, and final state effects. Our work shows that not only interactions within a single molecule define the band structure of thin film C$_{60}$ (as is the case in a molecular solid). Long range interactions between the molecules have a profound effect shaping the electronic structure of this material. Our results solve the missing link between electronic dispersions, vibronic loss, and the gas state spectra, paving the way for further investigations of the orbital character of the C$_{60}$ bands and the engineering of novel C$_{60}$ heterostructures, of interest for photovoltaic and optoelectronic applications.
\begin{acknowledgments}
{\ddag}~DWL and CO-A contributed equally to this work.

The Advanced Light Source and the Molecular Foundry (including computational resources) are supported by the Director, Office of Science, Office of Basic Energy Sciences, of the U.S.~Department of Energy (U.S.~DOE-BES) under contract No.~DE-AC02-05CH11231.   
DWL, AZ, and AL were supported by the sp$^2$ Program (KC2207) for ARPES measurements, and  SG and JN were supported by the Theory of Materials Program for DFT calculations, both funded by the U.S. DOE-BES, Materials Sciences and Engineering Division, under contract No.~DE-AC02-05CH11231.   
CO-A was funded by the U.S.~DOE-BES under contract DE-SC0018154 for sample fabrication, ARPES data acquisition and data analysis. CO-A would like to acknowledge fruitful discussions with V\'{e}ronique Brouet. SG acknowledges financial support by the Swiss National Science Foundation Early Postdoctoral Mobility Program. 
\end{acknowledgments}


%

\end{document}